\documentclass[onecolumn,showpacs,preprintnumbers,amsmath,amssymb]{revtex4-1}

\usepackage{graphicx}
\usepackage{dcolumn}
\usepackage{bm}
\usepackage{color}

\begin{document}

\preprint{}

\title{Weinberg operator contribution to the CP-odd nuclear force in the quark model}

\author{Nodoka Yamanaka$^{1,2}$}
\email{nyamanaka@kmi.nagoya-u.ac.jp}
\author{Makoto Oka$^{2,3}$}
\affiliation{$^1$Kobayashi-Maskawa Institute for the Origin of Particles and the Universe, Nagoya University, Furocho, Chikusa, Aichi 464-8602, Japan}
\affiliation{$^2$Nishina Center for Accelerator-Based Science, RIKEN, Wako 351-0198, Japan}
\affiliation{$^3$Advanced Science Research Center, Japan Atomic Energy Agency, Tokai, Ibaraki, 319-1195 Japan}

\date{\today}

\begin{abstract}
The contribution of the CP violating three-gluon interaction, proposed by Weinberg, to the short-range CP-odd nuclear force is evaluated in the nonrelativistic quark model.
We first show that the naive leading contribution generated by the quark exchange process vanishes at sufficiently short distance within the resonating group method, by considering the one-loop level gluon exchange CP-odd interquark potential induced by the Weinberg operator with massive quarks and gluons.
We then estimate the true leading contribution by evaluating the gluonic correction to the CP-odd interquark potential in the closure approximation.
It is found that the resulting irreducible CP-odd nuclear force is comparable to that generated by the chiral rotation of the CP-even short-range nuclear force, where the CP-odd mass calculated with QCD sum rules is used as input.
The explicit calculation of the electric dipole moment (EDM) of the $^3$He nucleus yields $d_{^3{\rm He}}^{\rm (irr)} (w) = - 1.5 \, w\, e \, {\rm MeV}$.
The total $^3$He EDM, accounting for the intrinsic nucleon EDM, the pion-exchange and the short-range CP-odd nuclear force, is $d_{^3{\rm He}}^{\rm (tot)} (w) = 20_{-11}^{+14} \, w\, e \, {\rm MeV}$, with the dominant effect coming from the intrinsic nucleon EDM.
\end{abstract}

\pacs{11.30.Er,12.39.Jh,21.30.-x,13.40.Em}

\maketitle

\section{Introduction}
\label{sec:intro}

The particles of the standard model (SM) were all discovered with the advent of the Higgs boson in LHC experiments \cite{Aad:2012tfa,Chatrchyan:2012ufa}, but there are still phenomena which cannot be explained within this framework such as the matter abundance of the Universe.
In this particular case, CP violation of the fundamental theory is needed to generate the matter excess over antimatter in the early era of the Universe \cite{Sakharov:1967dj}.
However, the observed baryon-to-photon ratio, which is the direct translation of the baryon number asymmetry of the early Universe, is much larger than the SM prediction \cite{Kobayashi:1973fv,Farrar:1993hn,Huet:1994jb}.
Some new physics beyond the SM containing large CP violation is therefore needed to cure this problem.

As an ideal experimental probe of the CP violation beyond the SM, we have the electric dipole moment (EDM) \cite{He:1990qa,Bernreuther:1990jx,khriplovichbook,Pospelov:2005pr,Fukuyama:2012np,Hewett:2012ns,Engel:2013lsa,Yamanaka:2014mda,Ginges:2003qt,Yamanaka:2017mef,Chupp:2017rkp,Alarcon:2022ero}.
In addition to its high sensitivity to new physics, it has other notable advantages such as the measurability in many systems (neutron, nuclei, atoms, muons, etc), the small SM background \cite{Kobayashi:1973fv,Seng:2014lea,Yamanaka:2015ncb,Yamanaka:2016fjj,Lee:2018flm,Yamaguchi:2020eub,Ema:2022yra}, experimental cost, etc.
There are already experimental results \cite{Muong-2:2008ebm,Graner:2016ses,Bishof:2016uqx,ACME:2018yjb,Allmendinger:2019jrk,Abel:2020gbr}, as well as plans which aim to probe new physics scales far beyond the reach of LHC experiment, assuming $O(1)$ couplings and CP phases \cite{Farley:2003wt,Sakemi:2011zz,Anastassopoulos:2015ura,Vutha:2017pej,Flambaum:2018kuh,Flambaum:2019wit,Omarov:2020kws,Zakharova:2020fam,Porshnev:2021xdz,Ready:2021frx,Koksal:2021gyd,Terrano:2021zyh,Flambaum:2022pov,bnl}.

\begin{figure}[htb]
\begin{center}
\includegraphics[width=4cm]{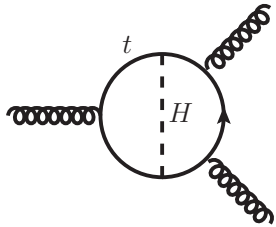}
\caption{\label{fig:weinberg_higgs}
Typical process contributing to the WO in SM with extended Higgs sector.
}
\end{center}
\end{figure}

To extract the CP violation of the quark-gluon sector, we need to theoretically analyze the intermediate hadron, nuclear and atomic level processes.
The most difficult task is the evaluation of the hadron matrix elements of quark-gluon level operators, due to the nonperturbative physics of QCD.
In the context of the study of the CP violation beyond the SM, the most interesting ones are the dimension-four, -five, and -six CP-odd operators which have the lowest mass dimension in the SM effective field theory \cite{Bonnefoy:2021tbt}.
Among them, the less studied one is the dimension-six CP-odd gluonic interaction (the so-called Weinberg operator, abbreviated as WO) \cite{Weinberg:1989dx}
\begin{eqnarray}
{\cal L}_W 
&=& 
\frac{1}{3!} w 
f^{abc} \epsilon^{\alpha \beta \gamma \delta} G^a_{\mu \alpha } G_{\beta \gamma}^b G_{\delta}^{\ \ \mu,c}
,
\label{eq:weinberg_operator}
\end{eqnarray}
where $a, b$ and $c$ are the color indices of the adjoint representation and the Greek alphabets are Lorentz indices, $f^{abc}$ is the structure constant of the $SU$(3) Lie algebra.
The contribution of this operator to the EDM is less known than the other ones since it does not involve quarks so that calculational techniques based on chiral effective field theory ($\chi$EFT) could not be used.
However, this does not mean that the WO is less important.
Indeed, it is sensitive to the extension of the SM Higgs sector which may contain additional CP phases (see Fig. \ref{fig:weinberg_higgs}) \cite{Weinberg:1989dx,Dicus:1989va,Boyd:1990bx,Cheng:1990gg,Bigi:1991rh,Bigi:1990kz,Hayashi:1994xf,Hayashi:1994ha,Wu:1994vx,Jung:2013hka,Brod:2013cka,Dekens:2014jka,Cirigliano:2016nyn,Cirigliano:2016njn,Panico:2017vlk,Cirigliano:2019vfc,Haisch:2019xyi,Cheung:2020ugr}.
It is also induced in many other candidates of new physics beyond the SM such as the supersymmetric models \cite{Dine:1990pf,Dai:1990xh,Arnowitt:1990je,Abel:2001vy,Demir:2002gg,Demir:2003js,Degrassi:2005zd,Abel:2005er,Ellis:2008zy,Li:2010ax,Zhao:2013gqa,Sala:2013osa,Hisano:2015rna,Nakai:2016atk,Yan:2020ocy}, and other specific models \cite{Chang:1990sfa,Rothstein:1990vd,Xu:2009nt,Choi:2016hro,Abe:2017sam,Dekens:2018bci,DiLuzio:2020oah,Dekens:2021bro,Gisbert:2021htg}.
In many models, the quark chromo-electric dipole moment first appears, and the WO is subleading (see e.g. Refs. \cite{Faessler:2006vi,Faessler:2006at}).
However, the WO becomes dominant when there are colored vectorlike quarks with CP violation which only interact with quarks via gluons, or when light axions are only coupled to heavy quarks via CP violating interactions.
In these important cases, it is mandatory to quantify the WO operator contribution to hadronic CP violation.
We note that the SM contribution is strongly suppressed by the GIM mechanism \cite{Booth:1992tv,Pospelov:1994uf,Yamaguchi:2020dsy}.

Let us now briefly review the evaluation of the WO contribution to observable EDMs, and then state which hadron level process has to be investigated for the quantification.
The effect of the WO on the neutron EDM was first estimated using the naive dimensional analysis when it was first introduced by Weinberg \cite{Weinberg:1989dx}.
Soon after this, it was pointed out that the neutron EDM has two leading hadron level processes \cite{Bigi:1990kz,Bigi:1991rh}, namely the reducible one generated by the chiral rotation of the neutron anomalous magnetic moment \cite{Porshnev:2021xdz,Baryshevsky:2020aez,Baryshevsky:2021kvc}, and the irreducible contact term.
The reducible contribution was then calculated using QCD sum rules \cite{Demir:2002gg,Haisch:2019bml}, and the irreducible one was evaluated within the quark model \cite{Yamanaka:2020kjo} (there also exist other hadron level estimations in the literature \cite{Chemtob:1991vv,Dib:2006hk}).
The total EDM of the nucleon is
\begin{eqnarray}
d_N (w) 
&\approx&
\left\{
\begin{array}{rl}
 (20\pm 12) \, w \, e \, {\rm MeV} & (N = n ) \cr
(- 18\pm 11)  \, w \, e \, {\rm MeV} & (N = p ) \cr
\end{array}
\right.
,
\label{eq:weinbergop_NEDM}
\end{eqnarray}
with an estimated uncertainty of 60\%.
Here $w$ is expressed in the unit of GeV$^{-2}$.
There are also other interesting approaches, such as the analysis of higher twist contribution of the parton distribution functions \cite{Hatta:2020ltd,Hatta:2020riw}, instanton models \cite{Weiss:2021kpt}, etc.
The most promising way is to use lattice QCD, but quantitative results are not yet available \cite{Abramczyk:2017oxr,Cirigliano:2020msr,Dragos:2019oxn,Rizik:2020naq}.
It is also interesting to note that the WO is flavor blind, and it may be probed with the EDM of other flavored baryons \cite{Baryshevsky:2020pbv,Aiola:2020yam,Biryukov:2021cml,Unal:2021lhb}.

\begin{figure}[htb]
\begin{center}
\includegraphics[width=14cm]{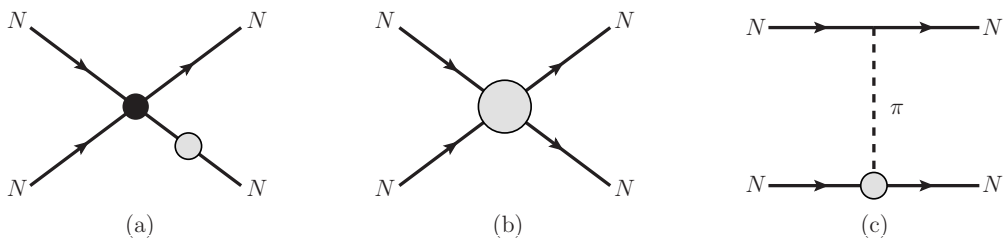}
\caption{\label{fig:CPV_nuclear_force}
Leading $\chi$EFT contribution to the CP-odd nuclear force induced by the WO.
Diagram (a) represents the chiral rotation of the CP-even nuclear force (black blob) by the CP-odd nucleon mass (gray blob).
Diagram (b) is the irreducible CP-odd nuclear force, evaluated in this work.
Diagram (c) is the one-pion exchange contribution, subleading in $\chi$EFT, but not negligible for heavy nuclei.
Permutations are not displayed.
}
\end{center}
\end{figure}

For the case of the nuclear and atomic EDMs, the CP-odd nuclear force also largely contributes, and we expect an enhancement of the CP violation due to the many-body effect.
It has also recently been pointed out that paramagnetic systems may also probe it via higher order electromagnetic interaction \cite{Flambaum:2020xcj}.
According to $\chi$EFT, the leading contribution of the WO to the CP-odd nuclear force is given by the short-range nucleon-nucleon interaction \cite{deVries:2011an,Bsaisou:2014zwa,Bsaisou:2014oka,Yamanaka:2016umw}
\begin{equation}
{\cal L}_C
=
-\bar C_1
m_N
\bar N N \, 
\bar N i \gamma_5 N
.
\label{eq:contactCPVNN}
\end{equation}
This contact interaction is converted to a CP-odd potential with a delta function in the coordinate space.
In the practical calculation, the delta function is smeared using $\lim_{\Lambda \to \infty} \Lambda^2 e^{-\Lambda r }/ 4 \pi r \to \delta (\vec{r})$.
By taking $\Lambda = m_\omega = 780$ MeV, Eq. (\ref{eq:contactCPVNN}) yields a short-range CP-odd nuclear force
\begin{eqnarray}
{\cal H}_{C} (\vec{R})
&=&
\frac{\bar C_1 m_\omega^3 }{8 \pi }
\Bigl(
\vec{\Sigma}_1 - \vec{\Sigma}_2
\Bigr) 
\cdot 
\vec{R } \,
\frac{e^{-m_\omega R}}{R^2}
\Biggl(
1 + \frac{1}{m_\omega R}
\Biggr)
\equiv
V_{C} (R)
\frac{\vec{R}}{R} \cdot \Bigl(\vec{\Sigma}_1- \vec{\Sigma}_2 \Bigr) 
,
\label{eq:vectormesonCPVNN}
\end{eqnarray}
where $\vec{\Sigma}_1$ and $ \vec{\Sigma}_2$ denote the spin Pauli matrices of the nucleons 1 and 2, respectively, and $\vec{R}$ is their relative coordinate oriented to the nucleon 1.
In this case too, we have the reducible process generated by the chiral rotation of the CP-even short-range nuclear force and the irreducible contact interaction [see Figs. \ref{fig:CPV_nuclear_force} (a) and (b)].
We also note that the one-pion exchange process [Fig. \ref{fig:CPV_nuclear_force} (c)], which is a priori subleading in $\chi$EFT, was found to not be negligible for heavy atoms through QCD sum rules calculation \cite{Osamura:2022rak}.

The reducible contribution may be evaluated like the nucleon EDM, by chirally rotating the CP-even contact interaction ${\cal L}_{NN} = -\frac{1}{2} C_S \bar N N \, \bar N N$ with $C_S = -120.8$ GeV$^{-2}$ obtained from $\chi$EFT analysis \cite{Epelbaum:2008ga}.
The contact CP-odd nuclear coupling is then
\begin{equation}
\bar C_1
=
\frac{m_{CP}}{m_N^2} C_S
,
\end{equation}
with the CP-odd nucleon mass $m_{CP}$ calculated using QCD sum rules \cite{Demir:2002gg,Haisch:2019bml}
\begin{equation}
m_{CP}
=
-
\langle N | 
{\cal L}_w 
| N \rangle
=
-m_N w \frac{3 g_s m_0^2}{32 \pi^2}
\ln \left( \frac{M^2}{\mu_{\rm IR}^2} \right)
,
\label{eq:CPVmass}
\end{equation}
where $m_0^2 = (0.8 \pm 0.2 )$ GeV$^2$ and $\frac{M}{\mu_{\rm IR}} = \sqrt{2} (1.5 \pm 0.5)$ \cite{Haisch:2019bml}.
We use the one-loop level running of the QCD coupling $g_s = 2.13 \pm 0.03$, renormalized at $\mu = 1$ GeV.

Let us now see concrete values of the nuclear EDM of $^3$He generated by the WO.
The measurement of the EDMs of the deuteron and the $^3$He nucleus is expected to be performed using storage rings \cite{bnl} with an impressive projected sensitivity of $O(10^{-29})e$ cm, so the quantification of the WO contribution to them is of importance.
Since the deuteron EDM does not receive contribution from the isoscalar short-range CP-odd nuclear force, we focus on that of $^3$He.
With the reducible short-range CP-odd nuclear force [see Fig. \ref{fig:CPV_nuclear_force} (a)], the EDM of $^3$He nucleus was estimated to be  \cite{Song:2012yh,Yamanaka:2015qfa,Yamanaka:2016umw,Froese:2021civ,Yamanaka:2021doa,Osamura:2022rak}
\begin{eqnarray}
d_{^3{\rm He}}^{\rm (red)} (w) 
&=&
2_{-1}^{+8} w \, e \, {\rm MeV}
,
\label{eq:HeEDM}
\end{eqnarray}
where the central value is the result of the calculation with the Argonne $v18$ nuclear force \cite{Wiringa:1994wb}, and the error bar is due to other realistic nuclear forces \cite{Song:2012yh,Yamanaka:2016umw} and also to the 50\% uncertainty of the QCD sum rules evaluation of $m_{CP}$ \cite{Haisch:2019bml}.
We also show the result of the QCD sum rules calculation of the CP-odd pion-exchange nuclear force induced by the WO [see Fig. \ref{fig:CPV_nuclear_force} (c)] \cite{Osamura:2022rak}:
\begin{eqnarray}
d_{^3{\rm He}}^{(\pi)} (w) 
&\approx&
\pm [0.4-2.3] w \, e \, {\rm MeV} 
,
\label{eq:pionexchangeHeEDM}
\end{eqnarray}
where the range indicates the theoretical uncertainty band due to the QCD sum rules.
We note that the sign is not determined.
The above contributions from the CP-odd nuclear force are smaller than the valence nucleon EDM contribution
\begin{eqnarray}
d_{^3{\rm He}}^{\rm (val)} (w) 
&\approx&
0.89\, d_n (w) 
-0.04\, d_p (w) 
=
( 19\pm 11)\, w \, e \, {\rm MeV} 
,
\label{eq:valenceHeEDM}
\end{eqnarray}
where the coefficients of the second equality were calculated with the Argonne $v18$ nuclear force \cite{Wiringa:1994wb,Song:2012yh,Yamanaka:2015qfa}.

The irreducible contribution [Fig. \ref{fig:CPV_nuclear_force} (b)], however, has never been evaluated, and it is currently introducing a substantial systematic uncertainty in the analysis of nuclear and atomic EDMs.
The aim of this paper is to calculate this effect, and to control the systematics of the WO contribution.
The chiral rotation involves a $\gamma_5$, so it is a mixing of the nonrelativistic and relativistic components of the Dirac spinor.
This may then be considered as a relativistic effect.
On the other hand, the irreducible effect has no such feature, and it is mainly due to the internal excitation of the nucleon, so a nonrelativistic framework may work well in its evaluation.

In this work, we use the nonrelativistic quark model to calculate the WO contribution to the irreducible CP-odd nuclear force.
The quark model \cite{GellMann:1964nj} is successful in describing the spectra of many baryons as well as mesons with heavy quarks \cite{Sakita:1964qq,Isgur:1979be,Capstick:2000qj,Roberts:2007ni,Karliner:2008sv,Yoshida:2015tia,Hiyama:2018ukv,Aaij:2015tga,Isgur:1978xj,Isgur:1978wd,Godfrey:1985xj,Capstick:1986ter}.
In the context of our study, the most interesting point is that the short-range baryon-baryon interaction is also quite well described by the exchange of nonrelativistic quarks \cite{Oka:1980ax,Oka:1981ri,Oka:1981rj,Oka:1982qa,Oka:1983ku,Morimatsu:1984nh,Morimatsu:1984rn,Oka:1986fr,Takeuchi:1989yz,Oka:1989ud,Takeuchi:1990qj,Oka:1990vx,Inoue:1995fs,Sasaki:1999vi,Oka:2000wj,Fujiwara:2006yh,Park:2019bsz}, and this mechanism shows good consistency with recent lattice QCD results \cite{Ishii:2006ec,Aoki:2009ji,Inoue:2010es,Gongyo:2017fjb}.
Since the irreducible contribution of the CP-odd nuclear force is also a short-range process, we expect it to be well described within the same quark model framework.

This paper is organized as follows.
In the next section, we define the quark model and the resonating group method (RGM) used in this work to derive the WO contribution to the short-distance CP-odd nuclear force.
It is first shown that the naive leading contribution generated by the quark exchange process vanishes at short distance, and we then estimate the true leading contribution by evaluating the gluonic correction to the CP-odd interquark potential in the closure approximation.
In Section \ref{sec:result}, we calculate the EDM of the $^3$He nucleus and compare with the contribution generated by the chiral rotation.
We finally conclude in Section \ref{sec:summary}.

\section{Quark model calculation of the CP-odd nuclear force induced by the Weinberg operator\label{sec:quarkmodel}}

\subsection{CP-odd interquark potential}

\begin{figure}[htb]
\begin{center}
\includegraphics[width=4cm]{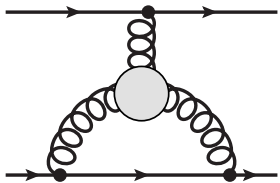}
\caption{\label{fig:2qint}
The interquark potential generated by the WO (gray blob) at the one-loop level.
Thick lines denote the nonrelativistic constituent quarks.
}
\end{center}
\end{figure}

Let us first show the CP-odd interquark force induced by the WO.
We use the CP-odd potential generated at the one-loop level (see Fig. \ref{fig:2qint}) as used in the calculation of the nucleon EDM \cite{Yamanaka:2020kjo}.
The explicit form is given by
\begin{eqnarray}
{\cal H}_{CPV,ij}
&=&
\frac{N_c g_s \alpha_s m_g^2 }{8 \pi }
w
(\vec{\sigma}_i- \vec{\sigma}_j) 
\cdot
\vec{\rho}_{ij}
\frac{e^{- m_g \rho_{ij} }}{\rho_{ij}^2}
\Biggl(
1+\frac{1}{ m_g \rho_{ij}}
\Biggr)
(t_a)_i \otimes (t_a)_j
\nonumber\\
&\equiv &
(\vec{\sigma}_i- \vec{\sigma}_j) 
\cdot \frac{\vec{\rho}_{ij}}{\rho_{ij}}
V(\rho_{ij})
(t_a)_i \otimes (t_a)_j
,
\label{eq:cpvhamiltonian}
\end{eqnarray}
where the subscripts $i,j$ denote the label of the interacting two nonrelativistic quarks, $\vec{\rho}_{ij}$ is their relative coordinate oriented to the $i$-th quark, $\vec{\sigma}_i$ the spin Pauli matrix and $(t_a)_i$ the generator of the $SU(3)_c$ group both acting on the $i$-th quark.
We note that the divergence due to the nonrenormalizability has been regularized using the heavy quark approximation.
The gluon mass $m_g\approx 350$ MeV was adopted respecting the result of Landau gauge lattice QCD \cite{Falcao:2020vyr}.
The strong coupling $\alpha_s \equiv \frac{g_s^2}{4\pi}$ was already given when we introduced the CP-odd mass in Eq. (\ref{eq:CPVmass}), but in this work we choose $\alpha_s = 0.5$.
If one wishes to calculate the WO contribution in some new physics model, its running down to the renormalization point $\mu =1$ GeV has to be taken into account \cite{Braaten:1990gq,Braaten:1990zt,Degrassi:2005zd,Hisano:2012cc,Dekens:2013zca,deVries:2019nsu,Kley:2021yhn,Degrande:2021zpv}.
We note that the WO (\ref{eq:weinberg_operator}) also generates the three-quark interaction in the constituent quark model, but its contribution is suppressed in the nonrelativistic approximation, so we neglect it \cite{Yamanaka:2020kjo}.

\subsection{CP-odd nuclear force from resonating group method}

Let us now present the formalism to derive the CP-odd nuclear force.
The interaction between two nucleons, each made of three nonrelativistic constituent quarks, may in principle be calculated with ab initio 6-body calculations of scattering, using e.g. the continuum discretized coupled channel method \cite{Hiyama:2003cu}.
However, the ab initio treatment is computationally very costly, and it is too demanding in view of the theoretical uncertainty of the quark model.
Here we propose to use the RGM \cite{Wheeler:1937zza,Wheeler:1937zz,Wildermuth,Shimodaya} which is a good analytical framework to handle the scattering of composite systems.
One of the most important feature of the RGM is that the forbidden states due to the antisymmetrization of constituent fermions may correctly be excluded, and the use of the harmonic oscillator basis is very practical.
The constituent quarks confined in the nucleon by the harmonic oscillator or linear potential have an almost Gaussian shape, so this fits well with the use of the RGM.
In fact, the repulsive core of the two-nucleon system was successfully derived within the above framework \cite{Oka:1980ax,Oka:1981ri,Oka:1981rj}.

\begin{figure}[htb]
\begin{center}
\includegraphics[width=8cm]{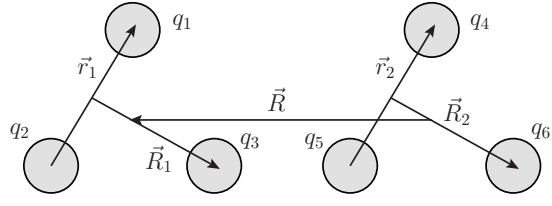}
\caption{\label{fig:NN_jacobi}
The Jacobi coordinates for the 2-nucleon system.
}
\end{center}
\end{figure}

The RGM equation to solve is 
\begin{equation}
\int
d \vec{\xi}_A
d \vec{\xi}_B
\psi_A^\dagger (\vec{\xi}_A)
\psi_B^\dagger (\vec{\xi}_B)
(H-E)
{\cal A}[
\psi_A (\vec{\xi}_A)
\psi_B (\vec{\xi}_B)
\chi_{AB}(\vec{R})
]
=0
,
\label{eq:RGM}
\end{equation}
where ${\cal A}$ is the antisymmetrization operator of the quarks inside the squared bracket.
Here $\vec{\xi}_{A,B}$ are the internal coordinates of the clusters (nucleons) $A,B$.
The Hamiltonian $H$ contains the quark model interactions, and in our case, the CP-odd interquark interaction ${\cal H}_{CPV}$ [see Eq. (\ref{eq:cpvhamiltonian})] is also present.
Here we may use the fact that the CP-odd nuclear force is small compared to the CP-even one, and also that it changes the parity.
The easiest way to analyze the scattering with this interaction is to consider the first order perturbation.
For that we just need to solve the CP-even RGM equation (\ref{eq:RGM}) separately for the two opposite parity states (we consider the lowest s- and p-waves) and then sandwich the CP-odd force (\ref{eq:cpvhamiltonian}) with them.
The final CP-odd internucleon interaction is then
\begin{equation}
H^r_{s p} (\vec{R} , \vec{R}')
=
\int d\vec{R}'' d\vec{R}'''
N_s^{-1/2} (\vec{R} , \vec{R}'')
H_{sp} (\vec{R}'' , \vec{R}''')
N_p^{-1/2} (\vec{R}''' , \vec{R}')
,
\label{CP-odd_nuclear_force}
\end{equation}
where 
\begin{eqnarray}
H_{s p} (\vec{R} , \vec{R}')
&=&
\sum_{i\ne j}
\int
d \vec{\xi}_A
d \vec{\xi}_B
d \vec{\Xi}
[\psi_A^\dagger (\vec{\xi}_A)
\psi_B^\dagger (\vec{\xi}_B)
]_s
\delta (\vec{R}-\vec{\Xi})
{\cal H}_{CPV,ij}
{\cal A}[
[\psi_A (\vec{\xi}_A)
\psi_B (\vec{\xi}_B)
]_p
\delta (\vec{R}'-\vec{\Xi})
]
,
\label{eq:RGMhamiltonian}
\\
N_{s / p} (\vec{R} , \vec{R}')
&=&
\int
d \vec{\xi}_A
d \vec{\xi}_B
d \vec{\Xi}
[\psi_A^\dagger (\vec{\xi}_A)
\psi_B^\dagger (\vec{\xi}_B)
]_{s,p}
\delta (\vec{R}-\vec{\Xi})
{\cal A}[
[\psi_A (\vec{\xi}_A)
\psi_B (\vec{\xi}_B)
]_{s,p}
\delta (\vec{R}'-\vec{\Xi})
]
,
\label{eq:RGMnorm}
\end{eqnarray}
Here $i,j$ run over all nonrelativistic quarks.
The above expressions depend on two external coordinates $\vec{R}$ and $\vec{R}'$ because the effective internucleon interaction is formally nonlocal.
In this subsection, we assume that the quarks are confined in nucleons with a common harmonic oscillator potential and we neglect all other CP-even interactions.
The s- and p-waves, between which the transition is considered, become therefore trivial. 
In this setup, it is convenient to use the Jacobi coordinates for $\vec{\xi}_A$ and $\vec{\xi}_B$ (see Fig. \ref{fig:NN_jacobi}) since the internal and the relative coordinates may be separated in the calculation so that we do not need to treat the center of mass motion explicitly.
We also use the local approximation $\vec{R} = \vec{R}'$ which should hold with nonrelativistic systems.

The single nucleon wave function in a nonrelativistic quark model is expressed as a direct product of color, spin, isospin, and radial components, entangled so as to antisymmetrize the state in the exchange of two arbitrary quarks.
The color part is by assumption a color SU(3) singlet, and a baryon state is completely antisymmetric.
In this work, we do not consider the internal excitation of the nucleon in the initial and final states, so the two internal Jacobi radial functions are both s-waves, as 
\begin{eqnarray}
\phi_i (\vec{\xi}_i)
=
\langle N | \vec{r}_i , \vec{R}_i \rangle
=
\left( 
\frac{2\omega}{\pi}
\right)^{\frac{3}{4}}
\left( 
\frac{2\Omega}{\pi}
\right)^{\frac{3}{4}}
e^{
-\omega {r}_i^2 
-\Omega {R}_i^2 
}
,
\end{eqnarray}
where $\omega = \frac{1}{4b^2}$ and $\Omega = \frac{1}{3b^2}$ are the oscillator constants with $b= 0.5$ fm.
The Jacobi coordinates of the single nucleon are defined in Fig. \ref{fig:NN_jacobi}.
Under this approximation, the spin-isospin part is 
\begin{eqnarray}
| {\rm spin} \rangle \otimes | {\rm isospin} \rangle 
&=&
\frac{1}{\sqrt{2}}
\sum_{\sigma =0,1}
|\, [\, [\chi_1 \otimes \chi_2]_\sigma \otimes \chi_3]_{1/2 , \Sigma_z} \rangle
\otimes
|\, [\, [\eta_1 \otimes \eta_2]_\sigma \otimes \eta_3]_{1/2 , T_z} \rangle
,
\label{eq:nucleonspin-isospin}
\end{eqnarray}
where $\chi_i$ and $\eta_i$ ($i=1,2,3$) are the spin and isospin states of the $i$-th quark.
Here $\Sigma_z$ and $T_z$ are the z-components of the spin and isospin of the single nucleon, respectively.
This form is derived by summing all three combinations of spin and isospin couplings respecting the antisymmetry of the state in the exchange of quarks (see Appendix E of Ref. \cite{Yamanaka:2014mda}).
As a consequence of this summation, the spin and isospin of the coupled two-quark system have the same quantum number $\sigma = 0$ or 1.
The two-nucleon wave function is a combination of two single nucleon ones as given above, and the interchange of the two nucleons is also antisymmetric, which restricts the total spin $S$, isospin $I$, and orbital angular momentum $L$ to $(-1)^{S+I+L} = -1$.

\begin{figure}[htb]
\begin{center}
\includegraphics[width=8cm]{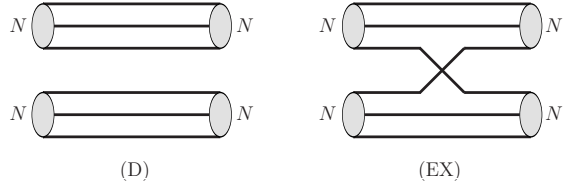}
\caption{\label{fig:norm_types}
Schematic diagrams of the norm kernel denoting the direct (D) and exchange (EX) contributions.
}
\end{center}
\end{figure}

Let us now calculate the two-nucleon matrix elements.
The norm kernel is given by
\begin{eqnarray}
N (\vec{R} , \vec{R}')
&=&
\langle
NN (\vec{R})
\, |\,
{\cal A}
\, |\,
NN (\vec{R}')
\rangle
=
C \langle
NN (\vec{R})
\, |\,
1-9P_{36}
\, |\,
NN (\vec{R}')
\rangle
\nonumber\\
&=&
C 
\Bigl[
\delta(\vec{R}-\vec{R}')
-9
\langle
NN (\vec{R})
\, |\,
P_{36}
\, |\,
NN (\vec{R}')
\rangle
\Bigr]
,
\end{eqnarray}
where $P_{36}$ denotes the exchange of quarks 3 and 6.
The factor 9 appeared due to the symmetry in the quark exchange, by also taking into account the fact that the exchange of two quarks between nucleons is equal to the exchange of single quarks times the exchange of nucleons which is also antisymmetric.
The total normalization $C$ contains the combinatoric factor arising from other quark exchanges which equally contribute to the Hamiltonian matrix elements.
Since this overall factor has no observable effects, we set $C=1$ from now on.
Processes contributing to the norm kernel are depicted in Fig. \ref{fig:norm_types}.

While the direct term is trivial, the exchange term requires some calculation.
The relevant color and spin-isospin matrix elements are listed in Table \ref{table:norm_exchange}.
For the calculation of the radial part, we use some approximation.
By using the expansion of the delta function in terms of the harmonic oscillator basis, the radial component of the two-nucleon state with fixed $\vec{R}$ may be rewritten as
\begin{equation}
|\, \phi_1(\vec{\xi}_1) \phi_2(\vec{\xi}_2) \delta(\vec{\Xi} - \vec{R}) \rangle
=
\sum_{nlm}
|\, \phi_1(\vec{\xi}_1) \phi_2(\vec{\xi}_2) \psi_{nlm} (\vec{\Xi}) \rangle \psi_{nlm}^\dagger (\vec{R})
\equiv
\sum_{nlm}
|\, nlm \rangle
\psi_{nlm}^\dagger (\vec{R})
,
\end{equation}
where $\psi_{nlm} (\vec{R}) $ is the harmonic oscillator eigenfunction.
The radial matrix element of the exchange norm kernel may then analytically be transformed as \cite{Takeuchi:1988bh}
\begin{eqnarray}
\langle
\vec{R}
\, |\,
P_{36}
\, |\,
\vec{R}'
\rangle
=
\sum_{nlmn'l'm'}
\langle
\vec{R}
\, |\,
nlm
\rangle
\langle
nlm 
\, |\,
P_{36}
\, |\,
n'l'm' 
\rangle
\langle
n'l'm'
\, |\,
\vec{R}'
\rangle
=
\sum_{nlm}
\psi_{nlm}^\dagger (\vec{R})
\frac{1}{3^{2n+l}}
\psi_{nlm} (\vec{R}')
.
\label{eq:norm_radial}
\end{eqnarray}
The exchange of quarks between two nucleons is a short-range process, so it is possible to approximate the above expression with the lowest lying eigenstates having $n=0$.
As we wrote above, we also take the local approximation $\vec{R}'=\vec{R}$ which will further simplify the calculation of matrix elements.

\begin{table}[htb]
\caption{
List of color and spin-isospin matrix elements of the quark exchange in two-nucleon system needed in the calculation of the norm kernel.
}
\begin{center}
\begin{tabular}{|l|cc|}
\hline
Type & Direct & Exchange \\ 
\hline
Color & 1 & 1/3 \\
\hline
Spin-isospin & & \\
$(S,I) = (0,0)$ & 1 & 7/9 \\
$(S,I) = (0,1)$ & 1 & -1/27 \\
$(S,I) = (1,0)$ & 1 & -1/27 \\
$(S,I) = (1,1)$ & 1 & 31/81 \\
\hline
\end{tabular}
\end{center}
\label{table:norm_exchange}
\end{table}

We explicit show below the forms of the s- and p-wave norm kernel with fixed $\vec{R}$ for the relevant channels:
\begin{eqnarray}
N_{L=0, S=1 ,I =0 } (\vec{R} , \vec{R}')
&\approx
&
\delta(\vec{R}-\vec{R}')
\times
4\pi 
\frac{10}{9}
\left( 
\frac{2\bar \Omega}{\pi}
\right)^\frac{3}{2}
e^{-2\bar \Omega R^2}
,
\\
N_{L=0, S=0 ,I =1 } (\vec{R} , \vec{R}')
&\approx
&
\delta(\vec{R}-\vec{R}')
\times
4\pi 
\frac{10}{9}
\left( 
\frac{2\bar \Omega}{\pi}
\right)^\frac{3}{2}
e^{-2\bar \Omega R^2}
,
\\
N_{L=1 , S=0 ,I =0 } (\vec{R} , \vec{R}')
&\approx
&
\delta(\vec{R}-\vec{R}')
\times
\frac{2}{9}
\frac{8}{3\sqrt{\pi}}
\left( 
2\bar \Omega
\right)^\frac{5}{2}
R^2
e^{-2\bar \Omega R^2}
,
\\
N_{L=1 , S=1 , J=0,I =1} (\vec{R} , \vec{R}')
&\approx
&
\delta(\vec{R}-\vec{R}')
\times
\frac{50}{81}
\frac{8}{3\sqrt{\pi}}
\left( 
2\bar \Omega
\right)^\frac{5}{2}
R^2
e^{-2\bar \Omega R^2}
,
\label{eq:normp2}
\end{eqnarray}
where we used $\psi_{00m} (\vec{R}) = \left( \frac{2\bar \Omega}{\pi} \right)^\frac{3}{4} e^{-\bar \Omega R^2} Y_{00}(\theta ,\phi)$ and $\psi_{1m} (\vec{R}) = \sqrt{\frac{8}{3\sqrt{\pi}}} \left( 2\bar \Omega \right)^\frac{5}{4} e^{-\bar \Omega R^2} R \, Y_{1m}(\theta ,\phi)$ with the oscillator constant $\bar \Omega = \frac{3}{4b^2}$.

\begin{figure}[htb]
\begin{center}
\includegraphics[width=18cm]{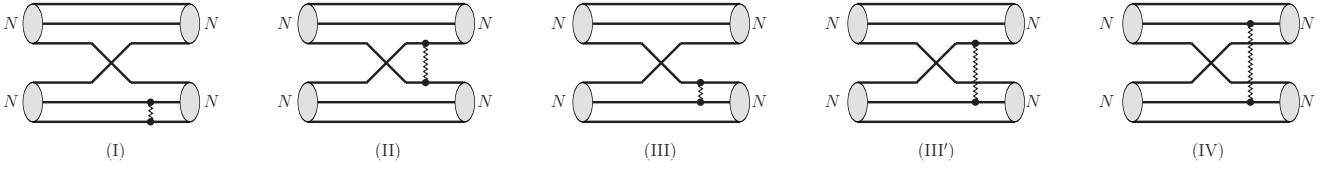}
\caption{\label{fig:potential_types}
Schematic diagrams of the potential kernel with several possibilities of quark exchange.
Jagged lines represent the CP-odd two-quark potential (\ref{eq:cpvhamiltonian}).
Type I does not contribute if we use the Jacobi coordinates because the internal excitation will change the nucleon to $N^*$.
}
\end{center}
\end{figure}

Let us now see the CP-odd potential kernel.
The CP-odd interquark potential (\ref{eq:cpvhamiltonian}) is a color exchanging interaction, so the apparent leading effect is given by quark exchange processes.
This antisymmetrization then brings five distinct processes as shown in Fig. \ref{fig:potential_types} \cite{Oka:1980ax,Oka:1981ri,Oka:1981rj}.
We note that Type I does not contribute if we use the Jacobi coordinates, since the CP-odd interaction between internal quarks which are not exchanged excites the nucleon.
We may explicitly write the matrix elements as
\begin{eqnarray}
&& \hspace{-2em}
\sum_{i\ne j}
\langle
NN (L=1, S=0 , I)
\, |\,
{\cal H}_{CPV,ij}
{\cal A}
\, |\,
NN (L'=0, S'=1 , I)
\rangle
\nonumber\\
&=&
-9\Bigl[
\langle
NN  (L=1, S=0 , I)
\, |\,
{\cal H}_{CPV,36}
P_{36}
\, |\,
NN (L'=0, S'=1 , I)
\rangle
\nonumber\\
&& \hspace{2em}
+4
\langle
NN  (L=1, S=0 , I)
\, |\,
{\cal H}_{CPV,45}
P_{35}
\, |\,
NN (L'=0, S'=1 , I)
\rangle
\nonumber\\
&& \hspace{2em}
+4
\langle
NN (L=1, S=0 , I)
\, |\,
{\cal H}_{CPV,36}
P_{16}
\, |\,
NN (L'=0, S'=1 , I)
\rangle
\nonumber\\
&& \hspace{2em}
+4
\langle
NN (L=1, S=0 , I)
\, |\,
{\cal H}_{CPV,36}
P_{14}
\, |\,
NN (L'=0, S'=1 , I)
\rangle
\Bigr]
,
\label{eq:potentialtypes}
\end{eqnarray}
where $\vec{\rho}_{ij}$ are the coordinates of the quarks.
We note that the spin structure of the CP-odd interquark interaction (\ref{eq:cpvhamiltonian}) is proportional to $\vec{\sigma}_i-\vec{\sigma}_j$, which forces the total spin of the two-nucleon system to change ($S' \to S$), while the orbital angular momentum also shifts by one unit ($L' \to L$), due to the CP-odd nature.
The isospin $I$ is conserved.

\begin{table}[htb]
\caption{
List of color and spin-isospin matrix elements of the quark exchange in a two-nucleon system needed in the calculation of the CP-odd potential kernel.
}
\begin{center}
\begin{tabular}{|l|cccc|}
\hline
Type & Type II & Type III & Type III' & Type IV \\ 
\hline
Color & 4/9 & -2/9 & -2/9  & 1/9 \\
\hline
Spin-isospin & & & & \\
$( S , S',I ) = (0,1,0)$ & -22/27 & 0 & -2/9 & 10/27 \\
$( S , S',I ) = (1,0,1)$ & -34/81 & -8/27 & -10/81 & 14/81 \\
$( S , S',I ) = (0,1,1)$ & 34/81 & 2/9 & 8/27 & 14/81 \\
$( S , S',I ) = (1,0,0)$ & 22/27 & 2/9 & 16/27 & 10/27 \\
\hline
\end{tabular}
\end{center}
\label{table:potential_exchange}
\end{table}

In Table \ref{table:potential_exchange}, we give the color and spin-isospin matrix elements of the potential kernel (the detailed derivation of the spin-isospin contribution is given in Appendix \ref{sec:spin-isospinME}).
Below, we give the analytic forms of the radial components for each contributing type:
\begin{eqnarray}
&&\hspace{-3em}
\langle
NN (L=1 , L_z=m)
\, |\,
[ V(\rho_{36}) \vec{\rho}_{36}/\rho_{36} ] P_{36}
\, |\,
NN (L'=0 , L'_z=0)
\rangle
\ \ \ \ \ \ \ \ ({\rm Type \, II})
\nonumber\\
&=&
\langle
NN (L=1 , L_z=m)
\, |\,
[ V(\rho_{36}) \vec{\rho}_{36}/\rho_{36}] P_{16}
\, |\,
NN (L'=0 , L'_z=0)
\rangle
\ \ \ \ \ \ \ \ ({\rm Type \, III'})
\nonumber\\
&=&
\langle
NN (L=1 , L_z=m)
\, |\,
[ V(\rho_{36}) \vec{\rho}_{36}/\rho_{36} ] P_{14}
\, |\,
NN (L'=0 , L'_z=0)
\rangle
\ \ \ \ \ \ \ \ ({\rm Type \, IV})
\nonumber\\
&=&
\sqrt{\frac{2}{3}}
\frac{8\bar \Omega^2}{\pi}
\left( 
\frac{2\Omega}{\pi}
\right)^3
e^{-2\bar \Omega R^2}
\int d\vec{R}_1 d\vec{R}_2 d\hat{R} \,
e^{
-2\Omega (R_1^2 + R_2^2)
}
\tilde V\left(
\left|
\frac{2}{3} ( \vec{R}_1 -\vec{R}_2 ) +\vec{R} 
\right|
\right)
\Bigl[
\frac{2}{3} ( \vec{R}_1 -\vec{R}_2 ) \cdot \vec{R} +R^2 
\Bigr]
\delta_{0m}
,
\\
&&\hspace{-3em}
\langle
NN (L=1 , L_z=m)
\, |\,
[ 
V(\rho_{45}) \vec{\rho}_{45}/\rho_{45}
] P_{35}
\, |\,
NN (L'=0 , L'_z=0)
\rangle
\ \ \ \ \ \ \ \ ({\rm Type \, III})
\nonumber\\
&=&
\sqrt{\frac{2}{3}}
\frac{8\bar \Omega^2}{\pi}
\left( 
\frac{2\omega}{\pi}
\right)^{\frac{3}{2}}
e^{
-2\bar \Omega R^2 
}
\int d\vec{r}_2 d\hat{R} \,
e^{
-2\omega r_2^2
}
\tilde V ( r_2 ) 
\Bigl[ 
\vec{r}_2 \cdot \vec{R}
\Bigr] 
\delta_{0m}
\nonumber\\
&=&
0
,
\end{eqnarray}
where $\tilde V(\rho) \equiv V(\rho) /\rho$ is defined using Eq. (\ref{eq:cpvhamiltonian}), and the $\hat{R}$ integral is the integral over the angular variables of $\vec{R}$.
Here we explicitly wrote in the parentheses the processes (as ``Type'') from which each matrix element arises (see Fig. \ref{fig:potential_types}).
We see that Type III vanishes and that the radial integrals are the same for Types II, III', and IV.
Owing to these facts, the total contribution for both $I=0,1$ channels vanishes [together with the combinatoric factors of Eq. (\ref{eq:potentialtypes}), the color, and spin-isospin matrix elements of Table \ref{table:potential_exchange}].

The cancellation found above is not accidental, but may be demonstrated as follows.
The CP-odd interquark potential (\ref{eq:cpvhamiltonian}) is strictly composed of color exchange terms, so we expect that it only works at short distance.
We then have a compact shell of 6-quark states, and the major part of the CP-odd potential matrix elements is given by the transition between the $(0{\rm s})^6$ and $(0{\rm s})^5(0{\rm p})$ states in the harmonic oscillator  basis.
Since the (0{\rm p}) state is proportional to $\vec{\rho} e^{-\rho^2 /b^2}$, we may factorize the matrix element as
\begin{eqnarray}
&&\hspace{-1em}
\langle
NN (L=1 , L_z=m)
\, |\,
[ 
V(\rho_{ij}) \vec{\rho}_{ij}/\rho_{ij}
] P_{km}
\, |\,
NN (L'=0 , L'_z=0)
\rangle
=
\vec{R}
\langle f (R) \rangle
,
\end{eqnarray}
where $i(k)$ and $j(m)$ label quarks which are in different nucleons. 
$\langle f (R) \rangle$ is some function of $R$, but does not depend on the choice of the quarks as long as quarks $i$ and $j$ belong to different nucleons.
We are now left with the calculation of the color-spin matrix elements
\begin{eqnarray}
&&
\langle
NN  (S , I)
\, |\,
(t_{ai})\otimes (t_{aj})
(\vec{\sigma}_i -\vec{\sigma}_j)
P_{km}
\, |\,
NN (S' , I)
\rangle
\nonumber\\
&=&
\langle
NN  (S , I)
\, |\,
[
(t_{ai} \vec{\sigma}_i )\otimes (t_{aj})
-
(t_{ai}) \otimes (t_{aj} \vec{\sigma}_j ) 
]
P_{km}
\, |\,
NN (S' , I)
\rangle
.
\label{eq:color-spin_factored_matrix}
\end{eqnarray}
By noting that the color generator $t_a$ has zero eigenvalue after summing over the three quarks that are inside a nucleon, $ \sum_{i =1,3} (t_{ai}) \, |\, N \rangle =0$, we may finally state that the above color-spin matrix elements, and consequently the WO matrix elements vanish.

\subsection{Estimation of the leading order contribution}

\begin{figure}[htb]
\begin{center}
\includegraphics[width=14cm]{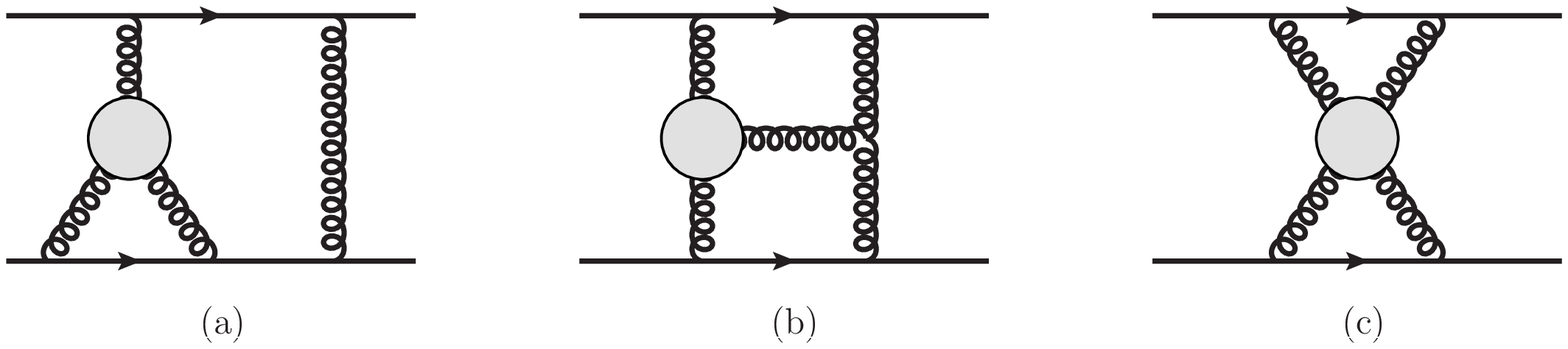}
\caption{\label{fig:CPVqq_2-loop}
Leading contribution to the color non-exchanging CP-odd potential between two nonrelativistic quarks (thick lines).
The wiggly lines denote the gluons.
The gray blob denotes the WO, which appears as a three-point vertex in the box diagram (a) as well as in diagram (b), and as a four-point vertex in diagram (c).
Permutations are not shown.
}
\end{center}
\end{figure}

In the previous section, we showed that the quark exchange contribution generated by the CP-odd interquark force (\ref{eq:cpvhamiltonian}) vanishes.
The leading contribution must then be a color non-exchanging process, and such effect should arise from the QCD correction to Eq. (\ref{eq:cpvhamiltonian}), generating a direct-type CP-odd nucleon-nucleon interaction.
With nonrelativistic quarks, we have three such processes, depicted in Fig. \ref{fig:CPVqq_2-loop}.
In this work, we focus on the box diagram [Fig. \ref{fig:CPVqq_2-loop} (a)] as the representative one, and we show that it may be matched with the closure approximation.

\begin{figure}[htb]
\begin{center}
\includegraphics[width=10cm]{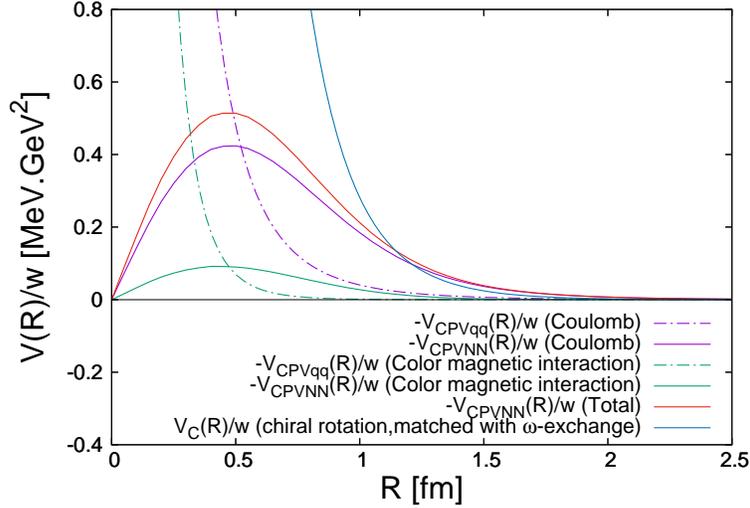}
\caption{\label{fig:result}
Result of the quark model calculation of the CP-odd nuclear force generated by the WO.
We plot separately the contribution from the Coulomb part and the color magnetic part of Eq. (\ref{eq:effectivecpvhamiltonian}).
The delta function of the color magnetic interaction was smeared using a Yukawa function with the smearing parameter $\Lambda = 600$ MeV which has the limit $\lim_{\Lambda \to \infty} \Lambda^2 e^{-\Lambda \rho }/ 4 \pi \rho \to \delta (\vec{\rho})$.
The contribution generated by the chiral rotation of the CP-even short-range nuclear force is also displayed for comparison [see Eq. (\ref{eq:contactCPVNN}) and below].
Note that we plotted $V_{CPVqq}$ and $V_{CPVNN}$ with a minus sign.
}
\end{center}
\end{figure}

The gluonic correction is given by the short-range one-gluon exchange interaction
\begin{equation}
{\cal H}_{g,ij}
=
\alpha_s
\Biggl[
\frac{1}{\rho_{ij}}
-
\frac{2 \pi}{3 m_Q^2 }
(\vec{\sigma}_i \cdot \vec{\sigma}_j) 
\delta ( \vec{\rho}_{ij} )
\Biggr]
(t_a)_i \otimes (t_a)_j
,
\end{equation}
where $m_Q \approx 300$ MeV is the mass of the nonrelativistic constituent quark.
The effective CP-odd potential for static quarks generated by Fig. \ref{fig:CPVqq_2-loop} (a) is then 
\begin{eqnarray}
{\cal H}_{CPVqq,ij} (\vec{\rho}_{ij})
&=&
-
\frac{N_c g_s \alpha_s^2 m_g^2 }{18 \pi \Lambda'}
w
\Biggl[
\frac{1}{\rho_{ij}}
(\vec{\sigma}_i- \vec{\sigma}_j) 
+
\frac{2 \pi}{3 m_Q^2 }
\delta (\vec{\rho_{ij}})
(\vec{\sigma}_i- \vec{\sigma}_j ) 
\Biggr]
\cdot
\vec{\rho}_{ij} \frac{e^{- m_g \rho_{ij} }}{ \rho_{ij}^2 + b_{\rm cut}^2} 
\Biggl(
1 + \frac{1}{m_g \rho_{ij}}
\Biggr)
\nonumber\\
&\equiv &
V_{CPVqq} (\rho)
\frac{\vec{\rho}}{\rho} \cdot (\vec{\sigma}_i- \vec{\sigma}_j) 
. 
\label{eq:effectivecpvhamiltonian}
\end{eqnarray}
Here we used the relations $(t_a)_i \otimes (t_a)_j \cdot (t_b)_i \otimes (t_b)_j = \frac{2}{9} 1_i \otimes 1_j + \cdots$ where the ellipses denote the color octet contribution, and $(\vec{\sigma}_i- \vec{\sigma}_j) (\vec{\sigma}_i \cdot \vec{\sigma}_j)  = \vec{\sigma}_j- \vec{\sigma}_i +2 i (\vec{\sigma}_j \times \vec{\sigma}_i) $.
We omitted the color octet part since it does not contribute to the direct as well as the exchange processes at short distance as shown in the previous section, and also the term with $2 i (\vec{\sigma}_j \times \vec{\sigma}_i)$ which cancels when the gluonic corrections acting before and after the CP-odd interaction (\ref{eq:cpvhamiltonian}) are summed up.
Here cutoff parameters $\Lambda' = 1$ GeV and $b_{\rm cut}=0.2$ fm were also introduced for convergence.
We will shortly see that $\Lambda'$ will be matched with the averaged energy of color octet intermediate nucleon states in the closure approximation.
We display in Fig. \ref{fig:result} the Coulomb and color magnetic parts [first and second terms of the square bracket of the first line of Eq. (\ref{eq:effectivecpvhamiltonian}), respectively].

Let us now see the two-nucleon matrix elements.
The CP-odd Hamiltonian matrix elements in the first order of perturbation are given by
\begin{equation}
\sum_{i,j}
\langle \psi' |
{\cal H}_{CPV,ij}
| \psi_0 \rangle
=
\sum_{i,j,k,l}
\Biggl[
\sum_{m\ne 0}
\frac{
\langle \psi' |
{\cal H}_{g,kl}
| \psi_m \rangle
\langle \psi_m |
{\cal H}_{CPV,ij}
| \psi_0 \rangle
}{E_0 - E_m}
+
\sum_{ m'\ne 0}
\frac{
\langle \psi' |
{\cal H}_{CPV,ij}
| \psi_{m'} \rangle
\langle \psi_{m'} |
{\cal H}_{g,kl}
| \psi_0 \rangle
}{E_0 - E_{m'}}
\Biggr]
,
\end{equation}
where $\psi_0$ and $\psi'$ are the lowest energy two-nucleon eigenstates of the CP-even Hamiltonian having each other opposite parities, and $m$ ($m'$) runs over all color octet CP-odd (-even) two-nucleon states.
The summations of ($i,j$) and ($k,l$) run each over quarks belonging to the nucleons (1,2).
The one-gluon exchange correction may act on all quarks of each nucleons.
However, when the gluons are attached to different quarks than those interacting through the CP-odd interquark force (\ref{eq:cpvhamiltonian}), the factorized $SU(3)$ generators summed over the three quarks of a single nucleon, $ \sum_{i =1,3} (t_{ai}) \, |\, N \rangle =0$, cancel, just like the case of Eq. (\ref{eq:color-spin_factored_matrix}).
We then have $i=k$ and $j=l$.
When we apply the closure approximation, moreover, the sum over intermediate states with color octet nucleons reduces to the identity operator.
The closure approximation of the above CP-odd two-nucleon matrix element may then be matched with Eq. (\ref{eq:effectivecpvhamiltonian}), as
\begin{eqnarray}
\sum_{i,j}
\langle \psi' |
{\cal H}_{CPV,ij}
| \psi_0 \rangle
&\approx&
-
\sum_{i,j}
\Biggl[
\frac{
\langle \psi' |
{\cal H}_{g,ij}
{\cal H}_{CPV,ij}
| \psi_0 \rangle
}{\langle \Delta E \rangle}
+
\frac{
\langle \psi' |
{\cal H}_{CPV,ij}
{\cal H}_{g,ij}
| \psi_0 \rangle
}{\langle \Delta E' \rangle}
\Biggr]
=
\sum_{i,j}
\langle \psi' |
{\cal H}_{CPVqq,ij}
| \psi_0 \rangle
,
\label{eq:closure}
\end{eqnarray}
where $\langle \Delta E \rangle \approx \langle \Delta E' \rangle \sim \Lambda' =1$ GeV is the averaged excitation energy in the transition from the singlet-singlet to octet-octet two-nucleon states.

Now let us calculate the CP-odd nuclear force according to Eq. (\ref{eq:RGMhamiltonian}).
Since the interquark potential of Eq. (\ref{eq:effectivecpvhamiltonian}) does not exchange quarks, this is just a double folding:
\begin{eqnarray}
{\cal H}_{CPVNN} (\vec{R})
&=&
\sum_{i,j}
\left( 
\frac{2\Omega}{\pi}
\right)^3
\int d\vec{R}_1 d\vec{R}_2  \,
e^{
-2\Omega (\vec{R}_1^2 + \vec{R}_2^2)
}
{\cal H}_{CPVqq,ij}
\left(
\frac{2}{3} ( \vec{R}_1 -\vec{R}_2 ) +\vec{R} 
\right)
\nonumber\\
&\equiv &
V_{CPVNN} (R)
\frac{\vec{R}}{R} \cdot \Bigl(\vec{\Sigma}_1- \vec{\Sigma}_2 \Bigr) 
.
\end{eqnarray}
In the nonrelativistic quark model, the spin matrix element is trivially one, as $\langle N | \vec{\sigma} | N \rangle = \vec{\Sigma}$.
We show the result of the integration in Fig. \ref{fig:result}, where the contributions of the Coulomb and color magnetic parts of Eq. (\ref{eq:effectivecpvhamiltonian}) are plotted separately.
By fitting with a simple functional, we obtain
\begin{eqnarray}
V_{CPVNN} (R)
&=&
\Bigl[
-
(1.63 \,{\rm MeV}/{\rm fm}) R e^{-R^2/(0.41\, {\rm fm^2})}
-
(0.18 \,{\rm MeV}/{\rm fm}) R e^{-R^2/(1.0\, {\rm fm}^2)}
\Bigr]
w \cdot {\rm GeV}^2
.
\label{eq:fittedCPVNN}
\end{eqnarray}
Here $R$ and $w$ are expressed in the unit of fm and GeV$^{-2}$, respectively.
Let us compare our result with the reducible contribution generated by the chiral rotation of the CP-even nuclear force [Fig. \ref{fig:CPV_nuclear_force} (a)].
We see that the irreducible CP-odd nuclear force and that generated by the chiral rotation of the short-range CP-even nuclear force have opposite sign, and comparable at $R \sim 1.5$ fm.

\section{EDM of $^3$He nucleus\label{sec:result}}

Now that we have the CP-odd nuclear force, we may evaluate the EDM of light nuclei \cite{Yamanaka:2015qfa,Yamanaka:2016umw,Yamanaka:2021doa,Yamanaka:2016itb,Yamanaka:2019vec}.
The nuclear EDM is defined as
\begin{eqnarray}
d_{A}
&=&
\sum_{i=1}^{A} \frac{e}{2} 
\langle \, \Psi_A \, |\, \tau_i^z \, {\cal R}_{iz} \, | \, \Psi_A \, \rangle
,
\label{eq:nucleonedmpolarization}
\end{eqnarray}
where $|\, \Psi_A \, \rangle$ is the polarized nuclear state along the $z$-axis, and $\vec{\cal R}_i$ are the coordinates of the $i$-th nucleon.
We calculate the EDM of the $^3$He nucleus which is expected to be measured in future storage ring experiments \cite{bnl}.
The $^3$He EDM may be evaluated exactly in the same way as Ref. \cite{Yamanaka:2015qfa} where the Gaussian expansion method \cite{Hiyama:2003cu} was used, by substituting the CP-odd potential by our result (\ref{eq:fittedCPVNN}).
We then obtain the following EDM
\begin{eqnarray}
d_{^3{\rm He}}^{\rm (irr)} (w) 
&=&
- 1.5 \, w\, e \, {\rm MeV} 
,
\label{eq:weinbergop_result}
\end{eqnarray}
where we used the Argonne $v18$ potential \cite{Wiringa:1994wb} for the CP-even nuclear force.
We also inspected the effect of the three-body force by adding a simple two-range Gaussian three-nucleon force fitted to reproduce the binding energy of $^3$He, but the difference was only 3\% (lowering the EDM), so we may safely neglect its effect.

Let us now analyze the total contribution.
In view of the fact that we only considered the first diagram of Fig. \ref{fig:CPVqq_2-loop}, and also by taking into account all the uncertainty of the quark model, it is fair to attribute a theoretical uncertainty of 100\% to our result (\ref{eq:weinbergop_result}).
The sum of the reducible and irreducible effects then becomes
\begin{eqnarray}
d_{^3{\rm He}}^{\rm (red+irr)} (w) 
&=&
0.5_{-2}^{+8} w \, e \, {\rm MeV}
,
\label{eq:redirrHeEDM}
\end{eqnarray}
where we added the error bar of Eq. (\ref{eq:HeEDM}) by quadrature.
We did not consider the pion-exchange contribution (\ref{eq:pionexchangeHeEDM}) for which the sign is not yet determined, but we may already say that the pion-exchange contribution is comparable to the short-range CP-odd nuclear force, even though the pion-exchange one is formally subleading in $\chi$EFT.
We then see that the intrinsic nucleon EDM contribution (\ref{eq:valenceHeEDM}) is dominant as regards the $^3$He EDM.
Gathering all, the final $^3$He EDM induced by the WO is
\begin{eqnarray}
d_{^3{\rm He}}^{\rm (tot)} (w) 
&=&
20_{-11}^{+14} w \, e \, {\rm MeV}
,
\label{eq:totalHeEDM}
\end{eqnarray}
where we added in quadrature the error bar from the short-range CP-odd nuclear force (\ref{eq:redirrHeEDM}), the pion-exchange effect (\ref{eq:pionexchangeHeEDM}) for which the upper bound was considered as the theoretical uncertainty, and the dominant valence nucleon (\ref{eq:valenceHeEDM}).

\section{Summary\label{sec:summary}}

In this paper, we discussed the contribution of the WO to the CP-odd nuclear force at short distance.
This contact interaction is the leading order effect in $\chi$EFT, and its unknown strength represented so far a substantial systematic ambiguity in the analysis of nuclear and atomic EDMs.
After its evaluation using the RGM within the nonrelativistic quark model, we found that the apparent leading quark exchange process vanishes at short distance.
We then estimated the gluonic correction to the CP-odd interquark potential in the closure approximation, and the resulting irreducible CP-odd nuclear force was comparable to the reducible one induced by the chiral rotation of the short-distance CP-even nucleon-nucleon interaction.
We then explicitly calculated the EDM of $^3$He nucleus which is expected to be measured in future storage ring experiment.
The sum of the reducible and irreducible short-distance CP-odd nuclear force contributions to the $^3$He EDM was found to be comparable to that of the pion-exchange generated by the WO, and much smaller than the intrinsic nucleon EDM effect.
We actually just terminated the first studies of the leading contribution of the WO and its systematics at the hadron level.

Now that we know that the contribution of the short-distance CP-odd nuclear force is small, let us enumerate the topics which have to be investigated to achieve the quantification of the WO effect.
The first work is to estimate the effect of the short-range CP-odd nuclear force to the CP-odd moments of nuclei.
Indeed, as we saw in Eq. (\ref{eq:HeEDM}), this contribution has a large uncertainty band, and we do not know whether it is really smaller than the intrinsic nucleon EDM as given in Eq. (\ref{eq:weinbergop_NEDM}).
For heavy nuclei, the one-pion exchange CP-odd nuclear force generated by the WO also becomes important, and it may in some cases be comparable to the nucleon EDM effect \cite{Osamura:2022rak,Yanase:2020agg,Yanase:2020oos}.
This process is also currently affected by a large uncertainty.
We expect it to be quantified by determining the sign of the pion-pole process generated by the WO, and also by fixing the problem of the inconsistency of the pion-nucleon sigma-term between phenomenology and lattice QCD \cite{Yamanaka:2018uud,Gupta:2021ahb,Courtoy:2022kca}.

\begin{acknowledgments}
The authors thank Emiko Hiyama for useful discussion and comments.
This work was supported by the Grant-in-Aid for Scientific Research (Grants No. 21H00132 and No. 20K03959) from the Japan Society For the Promotion of Science.
It was also partly supported by the US-Japan exchange program for studies of hadron structure and QCD (title of the project: Analysis of the hadron level effective interaction of the Weinberg operator).
The nuclear wave function was calculated using the code developed by Emiko Hiyama.
Numerical computation in this work was carried out at the Yukawa Institute Computer Facility.
\end{acknowledgments}

\appendix

\section{Derivation of the spin-isospin matrix elements of the CP-odd potential kernel\label{sec:spin-isospinME}}

We give below the explicit derivations of the spin-isospin matrix elements of the CP-odd potential kernel for each type.
We follow the convention of Eq. (\ref{eq:nucleonspin-isospin}).
We also use the modified 9-j symbol
\begin{eqnarray}
\chi
\left(
\begin{array}{ccc}
j_1 & j_2 & j_3 \cr
j_4 & j_5 & j_6 \cr
j_7 & j_8 & j_9 \cr
\end{array}
\right)
&\equiv&
\langle
[[j_1 \otimes j_2]_{j_3}\otimes [j_4 \otimes j_5]_{j_6}]_{j_9}
\, |\,
[[j_1 \otimes j_4]_{j_7}\otimes [j_2 \otimes j_5]_{j_8}]_{j_9}
\rangle
\nonumber\\
&=&
\sqrt{(2j_3 +1)(2j_6 +1)(2j_7 +1)(2j_8 +1)}
\left\{
\begin{array}{ccc}
j_1 & j_2 & j_3 \cr
j_4 & j_5 & j_6 \cr
j_7 & j_8 & j_9 \cr
\end{array}
\right\}
.
\end{eqnarray}

We first transform the spin-isospin matrix elements of Type II, Type III', and Type IV which depend on a common expression.

Type II:
\begin{eqnarray}
&&\hspace{-1em}
\langle NN (S,S_z ; I,I_z )
\, | \,
(\vec \sigma_{3z}
-
\vec \sigma_{6z}
)
P_{36}
\, | \,
NN ( S',S_z ; I,I_z ) \rangle 
\nonumber\\
&=&
\frac{1}{4}
\hspace{-2em}
\sum_{\sigma,{\sigma'},{\sigma''},{\sigma'''}=0,1}
\hspace{-2em}
\langle \,
[ [ [ \chi_1 \otimes \chi_2 ]_\sigma \otimes \chi_3 ]_{\frac{1}{2}}
\otimes 
[ [ \chi_4 \otimes \chi_5 ]_{\sigma'} \otimes \chi_6 ]_{\frac{1}{2}}
]_{S,S_z}
\, | \,
\vec \sigma_{3z}
-
\vec \sigma_{6z}
\, | \,
[ [ [ \chi_1 \otimes \chi_2 ]_{\sigma''} \otimes \chi_6 ]_{\frac{1}{2}}
\otimes 
[ [ \chi_4 \otimes \chi_5 ]_{\sigma'''} \otimes \chi_3 ]_{\frac{1}{2}}
]_{S',S_z}
\rangle
\nonumber\\
&&
\times
\langle \,
[ [ [ \eta_1 \otimes \eta_2 ]_\sigma \otimes \eta_3 ]_{\frac{1}{2}}
\otimes 
[ [ \eta_4 \otimes \eta_5 ]_{\sigma'} \otimes \eta_6 ]_{\frac{1}{2}}
]_{I,I_z}
\, | \,
1
\, | \,
[ [ [ \eta_1 \otimes \eta_2 ]_{\sigma''} \otimes \eta_6 ]_{\frac{1}{2}}
\otimes 
[ [ \eta_4 \otimes \eta_5 ]_{\sigma'''} \otimes \eta_3 ]_{\frac{1}{2}}
]_{I,I_z}
\rangle
\nonumber\\
&=&
\frac{1}{4}
\hspace{-2em}
\sum_{\sigma,{\sigma'}=0,1;
\Sigma , \Sigma'=1/2,3/2}
\chi
\left(
\begin{array}{ccc}
\sigma & \frac{1}{2} & \frac{1}{2} \cr
\frac{1}{2} & \sigma' & \frac{1}{2} \cr
\Sigma & \Sigma' & S' \cr
\end{array}
\right)
\chi
\left(
\begin{array}{ccc}
\sigma & \frac{1}{2} & \frac{1}{2} \cr
\frac{1}{2} & \sigma' & \frac{1}{2} \cr
\frac{1}{2} & \frac{1}{2} & I \cr
\end{array}
\right)
\times (-1)^{\frac{1}{2}-\Sigma'}
\nonumber\\
&&
\times
\langle \,
[ [ [ \chi_1 \otimes \chi_2 ]_\sigma \otimes \chi_3 ]_{\frac{1}{2}}
\otimes 
[ [ \chi_4 \otimes \chi_5 ]_{\sigma'} \otimes \chi_6 ]_{\frac{1}{2}}
]_{S,S_z}
\, | \,
\vec \sigma_{3z}
-
\vec \sigma_{6z}
\, | \,
[ [ [ \chi_1 \otimes \chi_2 ]_{\sigma} \otimes \chi_3 ]_{\Sigma}
\otimes 
[ [ \chi_4 \otimes \chi_5 ]_{\sigma'} \otimes \chi_6 ]_{\Sigma'}
]_{S',S_z}
\rangle
.
\end{eqnarray}

Type III':
\begin{eqnarray}
&&
\langle NN (S,S_z ; I,I_z )
\, | \,
(\vec \sigma_{z3}
-
\vec \sigma_{z6}
)
P_{16}
\, | \,
NN ( S',S_z ; I,I_z ) \rangle 
\nonumber\\
&=&
\frac{1}{4}
\hspace{-2em}
\sum_{\sigma,{\sigma'},{\sigma''},{\sigma'''}=0,1}
\hspace{-2em}
\langle \,
[ [ [ \chi_1 \otimes \chi_2 ]_\sigma \otimes \chi_3 ]_{\frac{1}{2}}
\otimes 
[ [ \chi_4 \otimes \chi_5 ]_{\sigma'} \otimes \chi_6 ]_{\frac{1}{2}}
]_{S,S_z}
\, | \,
\vec \sigma_{3z}
-
\vec \sigma_{6z}
\, | \,
[ [ [ \chi_6 \otimes \chi_2 ]_{\sigma''} \otimes \chi_3 ]_{\frac{1}{2}}
\otimes 
[ [ \chi_4 \otimes \chi_5 ]_{\sigma'''} \otimes \chi_1 ]_{\frac{1}{2}}
]_{S',S_z}
\rangle
\nonumber\\
&&
\times
\langle \,
[ [ [ \eta_1 \otimes \eta_2 ]_\sigma \otimes \eta_3 ]_{\frac{1}{2}}
\otimes 
[ [ \eta_4 \otimes \eta_5 ]_{\sigma'} \otimes \eta_6 ]_{\frac{1}{2}}
]_{I,I_z}
\, | \,
1
\, | \,
[ [ [ \eta_6 \otimes \eta_2 ]_{\sigma''} \otimes \eta_3 ]_{\frac{1}{2}}
\otimes 
[ [ \eta_4 \otimes \eta_5 ]_{\sigma'''} \otimes \eta_1 ]_{\frac{1}{2}}
]_{I,I_z}
\rangle
\nonumber\\
&=&
\frac{1}{4}
\hspace{-1.5em}
\sum_{\sigma,{\sigma'},\tau,{\sigma''},\tilde \tau =0,1; \Sigma, \Sigma' = 1/2,3/2}
\hspace{-4.5em}
(-1)^{\frac{1}{2}-\Sigma'}
\chi
\left(
\begin{array}{ccc}
\frac{1}{2} & \frac{1}{2} & \sigma'' \cr
\frac{1}{2} & 0 & \frac{1}{2} \cr
\tau & \frac{1}{2} & \frac{1}{2} \cr
\end{array}
\right)
\chi
\left(
\begin{array}{ccc}
\frac{1}{2} & \frac{1}{2} & \sigma'' \cr
\frac{1}{2} & 0 & \frac{1}{2} \cr
\tilde \tau & \frac{1}{2} & \frac{1}{2} \cr
\end{array}
\right)
\chi
\left(
\begin{array}{ccc}
\tau & \frac{1}{2} & \frac{1}{2} \cr
\frac{1}{2} & \sigma' & \frac{1}{2} \cr
\Sigma & \Sigma' & S' \cr
\end{array}
\right)
\chi
\left(
\begin{array}{ccc}
\tilde \tau & \frac{1}{2} & \frac{1}{2} \cr
\frac{1}{2} & \sigma' & \frac{1}{2} \cr
\frac{1}{2} & \frac{1}{2} & I \cr
\end{array}
\right)
\chi
\left(
\begin{array}{ccc}
\frac{1}{2} & \frac{1}{2} & \tau \cr
\frac{1}{2} & 0 & \frac{1}{2} \cr
\sigma & \frac{1}{2} & \Sigma \cr
\end{array}
\right)
\chi
\left(
\begin{array}{ccc}
\frac{1}{2} & \frac{1}{2} & \tilde \tau \cr
\frac{1}{2} & 0 & \frac{1}{2} \cr
\sigma & \frac{1}{2} & \frac{1}{2} \cr
\end{array}
\right)
\nonumber\\
&&
\times
\langle \,
[ [ [ \chi_1 \otimes \chi_2 ]_\sigma \otimes \chi_3 ]_{\frac{1}{2}}
\otimes 
[ [ \chi_4 \otimes \chi_5 ]_{\sigma'} \otimes \chi_6 ]_{\frac{1}{2}}
]_{S,S_z}
\, | \,
\vec \sigma_{3z}
-
\vec \sigma_{6z}
\, | \,
[ [ [ \chi_1 \otimes \chi_2 ]_{\sigma} \otimes \chi_3 ]_{\Sigma}
\otimes 
[ [ \chi_4 \otimes \chi_5 ]_{\sigma'} \otimes \chi_6 ]_{\Sigma'}
]_{S',S_z}
\rangle
.
\label{eq:typeIIIdashspinisospin}
\end{eqnarray}

Type IV:
\begin{eqnarray}
&&
\langle NN (S,S_z ; I,I_z )
\, | \,
(\vec \sigma_{3z}
-
\vec \sigma_{6z}
)
P_{14}
\, | \,
NN ( S',S_z ; I,I_z ) \rangle 
\nonumber\\
&=&
\frac{1}{4}
\hspace{-2em}
\sum_{\sigma,{\sigma'},{\sigma''},{\sigma'''}=0,1}
\hspace{-2em}
\langle \,
[ [ [ \chi_1 \otimes \chi_2 ]_\sigma \otimes \chi_3 ]_{\frac{1}{2}}
\otimes 
[ [ \chi_4 \otimes \chi_5 ]_{\sigma'} \otimes \chi_6 ]_{\frac{1}{2}}
]_{S,S_z}
\, | \,
\vec \sigma_{3z}
-
\vec \sigma_{6z}
\, | \,
[ [ [ \chi_4 \otimes \chi_2 ]_{\sigma''} \otimes \chi_3 ]_{\frac{1}{2}}
\otimes 
[ [ \chi_1 \otimes \chi_5 ]_{\sigma'''} \otimes \chi_6 ]_{\frac{1}{2}}
]_{S',S_z}
\rangle
\nonumber\\
&&
\times
\langle \,
[ [ [ \eta_1 \otimes \eta_2 ]_\sigma \otimes \eta_3 ]_{\frac{1}{2}}
\otimes 
[ [ \eta_4 \otimes \eta_5 ]_{\sigma'} \otimes \eta_6 ]_{\frac{1}{2}}
]_{I,I_z}
\, | \,
1
\, | \,
[ [ [ \eta_4 \otimes \eta_2 ]_{\sigma''} \otimes \eta_3 ]_{\frac{1}{2}}
\otimes 
[ [ \eta_1 \otimes \eta_5 ]_{\sigma'''} \otimes \eta_6 ]_{\frac{1}{2}}
]_{I,I_z}
\rangle
\nonumber\\
&=&
\frac{1}{4} \hspace{-1em}
\sum_{\sigma,{\sigma'},{\sigma''},{\sigma'''},\tau,\tau' =0,1}
\hspace{-1em}
\chi
\left(
\begin{array}{ccc}
\frac{1}{2} & \frac{1}{2} & \sigma'' \cr
\frac{1}{2} & 0 & \frac{1}{2} \cr
\tau & \frac{1}{2} & \frac{1}{2} \cr
\end{array}
\right)
\chi
\left(
\begin{array}{ccc}
\frac{1}{2} & \frac{1}{2} & \sigma'' \cr
\frac{1}{2} & 0 & \frac{1}{2} \cr
\tilde \tau & \frac{1}{2} & \frac{1}{2} \cr
\end{array}
\right)
\chi
\left(
\begin{array}{ccc}
0 & \frac{1}{2} & \frac{1}{2} \cr
\frac{1}{2} & \frac{1}{2} & \sigma''' \cr
\frac{1}{2} & \tau' & \frac{1}{2} \cr
\end{array}
\right)
\chi
\left(
\begin{array}{ccc}
0 & \frac{1}{2} & \frac{1}{2} \cr
\frac{1}{2} & \frac{1}{2} & \sigma''' \cr
\frac{1}{2} & \tilde \tau' & \frac{1}{2} \cr
\end{array}
\right)
\chi
\left(
\begin{array}{ccc}
\tau & \frac{1}{2} & \frac{1}{2} \cr
\frac{1}{2} & \tau' & \frac{1}{2} \cr
\Sigma & \Sigma' & S' \cr
\end{array}
\right)
\,
\chi
\left(
\begin{array}{ccc}
\tilde \tau & \frac{1}{2} & \frac{1}{2} \cr
\frac{1}{2} & \tilde \tau' & \frac{1}{2} \cr
\frac{1}{2} & \frac{1}{2} & I \cr
\end{array}
\right)
\nonumber\\
&&\times \,
\chi
\left(
\begin{array}{ccc}
\frac{1}{2} & \frac{1}{2} & \tau \cr
\frac{1}{2} & 0 & \frac{1}{2} \cr
\sigma & \frac{1}{2} & \Sigma \cr
\end{array}
\right)
\chi
\left(
\begin{array}{ccc}
\frac{1}{2} & \frac{1}{2} & \tilde \tau \cr
\frac{1}{2} & 0 & \frac{1}{2} \cr
\sigma & \frac{1}{2} & \frac{1}{2} \cr
\end{array}
\right)
\chi
\left(
\begin{array}{ccc}
0 & \frac{1}{2} & \frac{1}{2} \cr
\frac{1}{2} & \frac{1}{2} & \tau' \cr
\frac{1}{2} & \sigma' & \Sigma' \cr
\end{array}
\right)
\chi
\left(
\begin{array}{ccc}
0 &\frac{1}{2} & \frac{1}{2} \cr
\frac{1}{2} & \frac{1}{2} & \tilde \tau' \cr
\frac{1}{2} & \sigma' & \frac{1}{2} \cr
\end{array}
\right)
\times (-1)^{\frac{1}{2}-\Sigma'}
\nonumber\\
&&
\times
\langle \,
[ [ [ \chi_1 \otimes \chi_2 ]_\sigma \otimes \chi_3 ]_{\frac{1}{2}}
\otimes 
[ [ \chi_4 \otimes \chi_5 ]_{\sigma'} \otimes \chi_6 ]_{\frac{1}{2}}
]_{S,S_z}
\, | \,
\vec \sigma_{3z}
-
\vec \sigma_{6z}
\, | \,
[ [ [ \chi_1 \otimes \chi_2 ]_{\sigma} \otimes \chi_3 ]_{\Sigma}
\otimes 
[ [ \chi_4 \otimes \chi_5 ]_{\sigma'} \otimes \chi_6 ]_{\Sigma'}
]_{S',S_z}
\rangle
.
\end{eqnarray}

We see that Type II, Type III', and Type IV depend on a common matrix element
\begin{eqnarray}
&&
\langle \,
[ [ [ \chi_1 \otimes \chi_2 ]_\sigma \otimes \chi_3 ]_{\frac{1}{2}}
\otimes 
[ [ \chi_4 \otimes \chi_5 ]_{\sigma'} \otimes \chi_6 ]_{\frac{1}{2}}
]_{S,S_z}
\, | \,
\vec \sigma_{3z}
-
\vec \sigma_{6z}
\, | \,
[ [ [ \chi_1 \otimes \chi_2 ]_{\sigma} \otimes \chi_3 ]_{\Sigma}
\otimes 
[ [ \chi_4 \otimes \chi_5 ]_{\sigma'} \otimes \chi_6 ]_{\Sigma'}
]_{S',S_z}
\rangle
\nonumber\\
&=&
(-1)^{S-S_z}
\left(
\begin{array}{ccc}
S & 1 & S' \cr
-S_z & 0 & S_z \cr
\end{array}
\right)
\sqrt{\frac{2S+1}{2}}
\left[
\chi
\left(
\begin{array}{ccc}
\Sigma & \Sigma' & S'\cr
1 & 0 & 1 \cr
\frac{1}{2} & \frac{1}{2} & S \cr
\end{array}
\right)
\langle \,
 [ [ \chi_1 \otimes \chi_2 ]_\sigma \otimes \chi_3 ]_{\frac{1}{2}}
\, || \,
\vec \sigma_3 
\, || \,
 [ [ \chi_1 \otimes \chi_2 ]_{\sigma} \otimes \chi_3 ]_{\Sigma}
\rangle
\delta_{\Sigma' , \frac{1}{2}}
\right.
\nonumber\\
&& \hspace{16em}
\left.
-
\chi
\left(
\begin{array}{ccc}
\Sigma & \Sigma' & S'\cr
0 & 1 & 1 \cr
\frac{1}{2} & \frac{1}{2} & S \cr
\end{array}
\right)
\langle \,
[ [ \chi_4 \otimes \chi_5 ]_{\sigma'} \otimes \chi_6 ]_{\frac{1}{2}}
\, || \,
\vec \sigma_6
\, || \,
[ [ \chi_4 \otimes \chi_5 ]_{\sigma'} \otimes \chi_6 ]_{\Sigma'}
\rangle
\delta_{\Sigma , \frac{1}{2}}
\right]
. \ \ \ \ \ \ 
\end{eqnarray}

We also transform Type III which is irrelevant in this work:
\begin{eqnarray}
&&\hspace{-1em}
\langle NN (S,S_z ; I,I_z )
\, | \,
(\vec \sigma_{z4}
-
\vec \sigma_{z5}
)
P_{35}
\, | \,
NN ( S',S_z ; I,I_z ) \rangle 
\nonumber\\
&=&
\frac{1}{4}
\hspace{-2em}
\sum_{\sigma,{\sigma'},{\sigma''},{\sigma'''}=0,1}
\hspace{-2em}
\langle \,
[ [ [ \chi_1 \otimes \chi_2 ]_\sigma \otimes \chi_3 ]_{\frac{1}{2}}
\otimes 
[  [ \chi_4 \otimes \chi_5 ]_{\sigma'} \otimes \chi_6 ]_{\frac{1}{2}}
]_{S,S_z}
\, | \,
\vec \sigma_{4z}
-
\vec \sigma_{5z}
\, | \,
[ [ [ \chi_1 \otimes \chi_2 ]_{\sigma''} \otimes \chi_5 ]_{\frac{1}{2}}
\otimes 
[ [ \chi_4 \otimes \chi_3 ]_{\sigma'''} \otimes \chi_6 ]_{\frac{1}{2}}
]_{S',S_z}
\rangle
\nonumber\\
&&
\times
\langle \,
[ [ [ \eta_1 \otimes \eta_2 ]_\sigma \otimes \eta_3 ]_{\frac{1}{2}}
\otimes 
[ [ \eta_4 \otimes \eta_5 ]_{\sigma'} \otimes \eta_6 ]_{\frac{1}{2}}
]_{I,I_z}
\, | \,
1
\, | \,
[ [ [ \eta_1 \otimes \eta_2 ]_{\sigma''} \otimes \eta_5 ]_{\frac{1}{2}}
\otimes 
[ [ \eta_4 \otimes \eta_3 ]_{\sigma'''} \otimes \eta_6 ]_{\frac{1}{2}}
]_{I,I_z}
\rangle
\nonumber\\
&=&
\frac{1}{4} \hspace{-2em}
\sum_{\sigma,{\sigma'},\tau,{\sigma'''},\tilde \tau ,\lambda=0,1 ; \Sigma=1/2,3/2}
\hspace{-4.5em}
(-1)^{\frac{1}{2}-\Sigma}
\chi
\left(
\begin{array}{ccc}
0 & \frac{1}{2} & \frac{1}{2} \cr
\frac{1}{2} & \frac{1}{2} & \sigma''' \cr
\frac{1}{2} & \tau & \frac{1}{2} \cr
\end{array}
\right)
\chi
\left(
\begin{array}{ccc}
0 & \frac{1}{2} & \frac{1}{2} \cr
\frac{1}{2} & \frac{1}{2} & \sigma''' \cr
\frac{1}{2} & \tilde \tau & \frac{1}{2} \cr
\end{array}
\right)
\chi
\left(
\begin{array}{ccc}
\sigma & \frac{1}{2} & \frac{1}{2} \cr
\frac{1}{2} & \tau & \frac{1}{2} \cr
\frac{1}{2} & \Sigma & S' \cr
\end{array}
\right)
\chi
\left(
\begin{array}{ccc}
\sigma & \frac{1}{2} & \frac{1}{2} \cr
\frac{1}{2} & \tilde \tau & \frac{1}{2} \cr
\frac{1}{2} & \frac{1}{2} & I \cr
\end{array}
\right)
\chi
\left(
\begin{array}{ccc}
0 & \frac{1}{2} & \frac{1}{2} \cr
\frac{1}{2} & \frac{1}{2} & \tau \cr
\frac{1}{2} & \lambda & \Sigma \cr
\end{array}
\right)
\chi
\left(
\begin{array}{ccc}
0 & \frac{1}{2} & \frac{1}{2} \cr
\frac{1}{2} & \frac{1}{2} & \tilde \tau \cr
\frac{1}{2} & \sigma' & \frac{1}{2} \cr
\end{array}
\right)
\nonumber\\
&&
\times
\langle \,
[ [ [ \chi_1 \otimes \chi_2 ]_\sigma \otimes \chi_3 ]_{\frac{1}{2}}
\otimes 
[ [ \chi_4 \otimes \chi_5 ]_{\sigma'} \otimes \chi_6 ]_{\frac{1}{2}}
]_{S,S_z}
\, | \,
\vec \sigma_{4z}
-
\vec \sigma_{5z}
\, | \,
[ [ [ \chi_1 \otimes \chi_2 ]_{\sigma} \otimes \chi_3 ]_{\frac{1}{2}}
\otimes 
[ [ \chi_4 \otimes \chi_5 ]_{\lambda} \otimes \chi_6 ]_{\Sigma}
]_{S',S_z}
\rangle
\nonumber\\
&=&
\frac{1}{4}
\hspace{-2em}
\sum_{\sigma,{\sigma'},\tau,{\sigma'''},\tilde \tau ,\lambda=0,1 ; \Sigma=1/2,3/2}
\hspace{-4.5em}
(-1)^{\frac{1}{2}-\Sigma}
\chi
\left(
\begin{array}{ccc}
0 & \frac{1}{2} & \frac{1}{2} \cr
\frac{1}{2} & \frac{1}{2} & \sigma''' \cr
\frac{1}{2} & \tau & \frac{1}{2} \cr
\end{array}
\right)
\chi
\left(
\begin{array}{ccc}
0 & \frac{1}{2} & \frac{1}{2} \cr
\frac{1}{2} & \frac{1}{2} & \sigma''' \cr
\frac{1}{2} & \tilde \tau & \frac{1}{2} \cr
\end{array}
\right)
\chi
\left(
\begin{array}{ccc}
\sigma & \frac{1}{2} & \frac{1}{2} \cr
\frac{1}{2} & \tau & \frac{1}{2} \cr
\frac{1}{2} & \Sigma & S' \cr
\end{array}
\right)
\chi
\left(
\begin{array}{ccc}
\sigma & \frac{1}{2} & \frac{1}{2} \cr
\frac{1}{2} & \tilde \tau & \frac{1}{2} \cr
\frac{1}{2} & \frac{1}{2} & I \cr
\end{array}
\right)
\chi
\left(
\begin{array}{ccc}
0 & \frac{1}{2} & \frac{1}{2} \cr
\frac{1}{2} & \frac{1}{2} & \tau \cr
\frac{1}{2} & \lambda & \Sigma \cr
\end{array}
\right)
\chi
\left(
\begin{array}{ccc}
0 & \frac{1}{2} & \frac{1}{2} \cr
\frac{1}{2} & \frac{1}{2} & \tilde \tau \cr
\frac{1}{2} & \sigma' & \frac{1}{2} \cr
\end{array}
\right)
\nonumber\\
&&
\times 
(-1)^{S-S_z}
\left(
\begin{array}{ccc}
S & 1 & S' \cr
-S_z & 0 & S_z \cr
\end{array}
\right)
\sqrt{\frac{2S+1}{2}}
\chi
\left(
\begin{array}{ccc}
\frac{1}{2} & \frac{1}{2} & S'\cr
0 & 1 & 1 \cr
\frac{1}{2} & \frac{1}{2} & S \cr
\end{array}
\right)
\langle \,
[ [ \chi_4 \otimes \chi_5 ]_{\sigma'} \otimes \chi_6 ]_{\frac{1}{2}}
\, || \,
\vec \sigma_4
-
\vec \sigma_5
\, || \,
[ [ \chi_4 \otimes \chi_5 ]_{\lambda} \otimes \chi_6 ]_{\Sigma}
\rangle
.
\label{eq:typeIIIspinisospin}
\end{eqnarray}

We may further use the following formulae to derive the final matrix elements of Table \ref{table:potential_exchange}:
\begin{eqnarray}
\langle \,
 [ [ \chi_i \otimes \chi_j ]_\sigma \otimes \chi_k ]_{\frac{1}{2}}
\, || \,
1
\, || \,
 [ [ \chi_i \otimes \chi_j ]_{\sigma} \otimes \chi_k ]_{\frac{1}{2}}
\rangle
=
\langle \,
[ \chi_k \otimes [ \chi_i \otimes \chi_j ]_{\sigma'} ]_{\frac{1}{2}}
\, || \,
1
\, || \,
[ \chi_k \otimes [ \chi_i \otimes \chi_j ]_{\sigma'} ]_{\frac{1}{2}}
\rangle
&=&
\sqrt{2}
, \ \ \ \ \ \ \ 
\\
\langle \,
 [ [ \chi_i \otimes \chi_j ]_0 \otimes \chi_k ]_{\frac{1}{2}}
\, || \,
\vec \sigma_k
\, || \,
 [ [ \chi_i \otimes \chi_j ]_0 \otimes \chi_k ]_{\frac{1}{2}}
\rangle
=
\langle \,
[ \chi_k \otimes [ \chi_i \otimes \chi_j ]_0 ]_{\frac{1}{2}}
\, || \,
\vec \sigma_k
\, || \,
[ \chi_k \otimes [ \chi_i \otimes \chi_j ]_0 ]_{\frac{1}{2}}
\rangle
&=&
\sqrt{6}
,
\label{eq:yamanakabook1}
\\
\langle \,
 [ [ \chi_i \otimes \chi_j ]_1 \otimes \chi_k ]_{\frac{1}{2}}
\, || \,
\vec \sigma_k
\, || \,
 [ [ \chi_i \otimes \chi_j ]_1 \otimes \chi_k ]_{\frac{1}{2}}
\rangle
=
\langle \,
[ \chi_k \otimes [ \chi_i \otimes \chi_j ]_1 ]_{\frac{1}{2}}
\, || \,
\vec \sigma_k
\, || \,
[ \chi_k \otimes [ \chi_i \otimes \chi_j ]_1 ]_{\frac{1}{2}}
\rangle
&=&
-\frac{\sqrt{6}}{3}
,
\\
\langle \,
 [ [ \chi_i \otimes \chi_j ]_1 \otimes \chi_k ]_{\frac{1}{2}}
\, || \,
\vec \sigma_k
\, || \,
 [ [ \chi_i \otimes \chi_j ]_1 \otimes \chi_k ]_{\frac{3}{2}}
\rangle
=
-
\langle \,
[ \chi_k \otimes [ \chi_i \otimes \chi_j ]_1 ]_{\frac{1}{2}}
\, || \,
\vec \sigma_k
\, || \,
[ \chi_k \otimes [ \chi_i \otimes \chi_j ]_1 ]_{\frac{3}{2}}
\rangle
&=&
\frac{4\sqrt{3}}{3}
,
\\
\langle \,
 [ [ \chi_i \otimes \chi_j ]_0 \otimes \chi_k ]_{\frac{1}{2}}
\, || \,
\vec \sigma_{i}
\, || \,
 [ [ \chi_i \otimes \chi_j ]_0 \otimes \chi_k ]_{\frac{1}{2}}
\rangle
=
\langle \,
[ \chi_k \otimes [ \chi_i \otimes \chi_j ]_0 ]_{\frac{1}{2}}
\, || \,
\vec \sigma_{i}
\, || \,
[ \chi_k \otimes [ \chi_i \otimes \chi_j ]_0 ]_{\frac{1}{2}}
\rangle
&=&
0
\label{eq:nucleonspin_sigma45_0i}
,
\\
\langle \,
 [ [ \chi_i \otimes \chi_j ]_0 \otimes \chi_k ]_{\frac{1}{2}}
\, || \,
\vec \sigma_{j}
\, || \,
 [ [ \chi_i \otimes \chi_j ]_0 \otimes \chi_k ]_{\frac{1}{2}}
\rangle
=
\langle \,
[ \chi_k \otimes [ \chi_i \otimes \chi_j ]_0 ]_{\frac{1}{2}}
\, || \,
\vec \sigma_{j}
\, || \,
[ \chi_k \otimes [ \chi_i \otimes \chi_j ]_0 ]_{\frac{1}{2}}
\rangle
&=&
0
\label{eq:nucleonspin_sigma45_0j}
,
\\
\langle \,
 [ [ \chi_i \otimes \chi_j ]_1 \otimes \chi_k ]_{\frac{1}{2}}
\, || \,
\vec \sigma_{i}
\, || \,
 [ [ \chi_i \otimes \chi_j ]_1 \otimes \chi_k ]_{\frac{1}{2}}
\rangle
=
\langle \,
[ \chi_k \otimes [ \chi_i \otimes \chi_j ]_1 ]_{\frac{1}{2}}
\, || \,
\vec \sigma_{i}
\, || \,
[ \chi_k \otimes [ \chi_i \otimes \chi_j ]_1 ]_{\frac{1}{2}}
\rangle
&=&
2\frac{\sqrt{6}}{3}
,
\label{eq:nucleonspin_sigma45_1i}
\\
\langle \,
 [ [ \chi_i \otimes \chi_j ]_1 \otimes \chi_k ]_{\frac{1}{2}}
\, || \,
\vec \sigma_{j}
\, || \,
 [ [ \chi_i \otimes \chi_j ]_1 \otimes \chi_k ]_{\frac{1}{2}}
\rangle
=
\langle \,
[ \chi_k \otimes [ \chi_i \otimes \chi_j ]_1 ]_{\frac{1}{2}}
\, || \,
\vec \sigma_{j}
\, || \,
[ \chi_k \otimes [ \chi_i \otimes \chi_j ]_1 ]_{\frac{1}{2}}
\rangle
&=&
2\frac{\sqrt{6}}{3}
\label{eq:nucleonspin_sigma45_1j}
, 
\ \ \ \ \ \ \ 
\label{eq:yamanakabook2}
\end{eqnarray}
where $i \ne j \ne k \ne i$.

\end{document}